\documentclass[onecolumn,10pt]{article}
\usepackage{amsmath}
\usepackage{amsfonts}
\usepackage{amssymb}
\usepackage{amsthm}
\usepackage{times}
\usepackage{cancel}
\usepackage[dvips]{graphics,color}

\usepackage{tikz}

\title{\textbf{Why the Tsirelson Bound?}}
\author{Jeffrey Bub\\ \small \textit{Philosophy Department and Institute for Physical Science and Technology}\\  \small \textit{University of Maryland, College Park, MD 20742, USA}}
\date{}
\normalsize
\begin{document}

\maketitle

\begin{abstract}
Wheeler's question `why the quantum' has two aspects: why is the world quantum and not classical, and why is it quantum rather than superquantum, i.e., why the Tsirelson bound for quantum correlations? I discuss a remarkable answer to this question proposed by Paw\l owski et al \cite{Pawlowski+2009}, who provide an information-theoretic derivation of the Tsirelson bound from a principle they call `information causality.'
\end{abstract}

\bigskip

\section{Introduction}
In a remarkable information-theoretic derivation of the Tsirelson bound for quantum correlations by Paw\l owski et al \cite{Pawlowski+2009}, the authors derive the bound from a principle they call `information causality.' Here I review the original derivation and the information-theoretic principle involved, and consider the significance of the result. 

Einstein's special theory of relativity follows from just two principles: the light postulate and the principle of relativity. In a seminal paper \cite{PopescuRohrlich94}, Popescu and Rohrlich asked whether quantum mechanics follows from relativistic causality, the principle that causal processes or signals cannot propagate outside the light cone,  and nonlocality in the sense of Bell's theorem \cite{BellEPR}. They showed that it does not: quantum mechanics is only one of a class of theories consistent with these two principles. 

To see this, consider a `nonlocal box,' a hypothetical device proposed by Popescu and Rohrlich, now called a `Popescu-Rohrlich box' or PR-box. A PR-box has two inputs, $a \in \{0,1\}$ and $b\in \{0,1\}$, and two outputs, $A\in \{0,1\}$ and $B\in \{0,1\}$,\footnote{In a simulation of PR-box correlations by classical or quantum correlations,  inputs correspond to observables measured and outputs to measurement outcomes represented by real numbers, so it might seem more appropriate to use $A,B$ for inputs and $a,b$ for outputs. I follow the notation of Paw\l owski et al \cite{Pawlowski+2009} here, since this is the result I discuss in detail below. }  and is defined by the following correlations between inputs and outputs:
\begin{equation}
A\oplus B = a\cdot b \label{eqn:PRbox}
\end{equation}
where $\oplus$ is addition mod 2, i.e., 
\begin{itemize}
\item[(i)] same outputs (i.e., 00 or 11) if the inputs are 00 or 01 or 10 
\item[(ii)] different outputs (i.e., 01 or 10) if  the inputs are 11
\end{itemize}
together with a `no signaling' constraint.

A PR-box is bipartite and nonlocal in the sense that the $a$-input and $A$-ouput can be separated from the $b$-input and $B$-output by any distance without altering the correlations. For convenience, we can think of the  $a$-input as controlled by Alice, who monitors the $A$-output, and the $b$-input as controlled by Bob, who monitors the $B$-output. If we want the correlations of a PR-box to be consistent with relativistic causality, they should satisfy a `no signaling' constraint: no information should be available in the marginal probabilities of Alice's outputs about alternative input choices made by Bob, and conversely, i.e.,
\begin{eqnarray}
\sum_{b\in\{0,1\}}p(A,B|a,b)  =  p(A|a), \, A, a, b \in\{0,1\} \\
\sum_{a\in\{0,1\}}p(A,B|a,b)   =  p(B|b), \, B, a, b \in\{0,1\}
\end{eqnarray}
Note that `no signaling' is not a relativistic constraint \emph{per se}--it is simply a constraint on the marginal probabilities. But if this constraint is not satisfied, instantaneous (hence superluminal) signaling is possible, i.e., `no signaling' is a necessary condition for relativistic causality. 

It follows from (\ref{eqn:PRbox}) and `no signaling' that the  correlations are as in Table 1: 
 \begin{table}[h!]
\begin{center}
\begin{tabular}{|ll||ll|ll|} \hline
   &$a$&$0$ & &$1$&\\
   $b$&&&&&\\\hline\hline
  $0$ &&$p(00|00) = 1/2$&$ p(10|00) = 0$  & $p(00|10) = 1/2$&$ p(10|10) = 0$     \\
   &&$p(01|00) = 0$&$p(11|00) = 1/2$  & $p(01|10)=0$&$ p(11|10) = 1/2$  \\\hline
   $1$&&$p(00|01)=1/2$&$ p(10|01)=0$  & $p(00|11)=0$&$ p(10|11)=1/2$   \\
  &&$p(01|01)=0$&$ p(11|01)=1/2$  & $p(01|11)=1/2$&$ p(11|11)=0$   \\\hline
\end{tabular}
\end{center}
 \caption{PR-box correlations}
\end{table}

The probability $p(00|00)$ is to be read as $p(A=0,B=0|a=0,b=0)$, and the probability $p(01|10)$ is to be read as $p(A=0,B=1|a=1,b=0)$, etc. (I drop the commas for ease of reading; the first two slots in $p(--|--)$ before the conditionalization sign `$|$' represent the two possible outputs for Alice and Bob, respectively, and the second two slots after the conditionalization sign represent the two possible inputs for Alice and Bob, respectively.) Note that the sum of the probabilities in each square cell of the array in Table 1 is 1, and that the marginal probability of 0 for Alice or for Bob is obtained by adding the probabilities in the left column of each cell or the top row of each cell, respectively, and the marginal probability of 1 is obtained for Alice or for Bob by adding the probabilities in the right column of each cell or the bottom row of each cell, respectively. One could define a PR-box as exhibiting the correlations in Table 1, which are `no signaling,' rather than in terms of the condition $A\oplus B = a\cdot b$ and the `no signaling' constraint. 

Note that a PR box functions in such a way that if Alice inputs a 0 or a 1, her output is 0 or 1 with probability 1/2, irrespective of Bob's input, and irrespective of whether Bob inputs anything at all. Similarly for Bob. The requirement is simply that whenever there are in fact two inputs, the inputs and outputs are correlated according to (\ref{eqn:PRbox}). A PR-box can function only once, so to get the statistics for many pairs of inputs one has to use many PR-boxes. This avoids the problem of selecting the `corresponding' input pairs for different inputs at various times, which would depend on the reference frame. In this respect, a PR-box is like a quantum system: after a system has responded to a measurement (produced an output for an  input), the system is no longer in the same quantum state, and one has to use many systems prepared in the same quantum state to exhibit the probabilities associated with a given quantum state.

What is the optimal probability that Alice and Bob can simulate a PR-box, supposing they are allowed certain resources? 

In units where $A = \pm 1, B = \pm 1$,\footnote{It is convenient to change units here to relate the probability to the usual expression for the Clauser-Horne-Shimony-Holt correlation, where the expectation values are expressed in terms of $\pm 1$ values for $A$ and $B$ (the relevant observables). Note that `same output' or `different output' mean the same thing whatever the units, so the probabilities $p(\mbox{same output}|AB)$ and $p(\mbox{different output}|AB)$ take the same values whatever the units, but the expectation value $\langle AB\rangle$ depends on the units for $A$ and $B$.}
\begin{equation}
\langle 00\rangle = p(\mbox{same output}|00) - p(\mbox{different output}|00)
\end{equation}
so:
\begin{eqnarray}
p(\mbox{same output}|00) & = & \frac{1 +  \langle 00\rangle}{2} \\
p(\mbox{different output}|00) & = & \frac{1-\langle 00\rangle}{2}
\end{eqnarray}
and similarly for input pairs 01, 10, 11. 

It follows that the probability of successfully simulating a PR-box is given by:
\begin{eqnarray}
\mbox{p(successful sim)} & = & \frac{1}{4}(p(\mbox{same output}|00) + p(\mbox{same output}|01) + \nonumber \\
& &  p(\mbox{same output}|10) + p(\mbox{different output}|11)) \\
& = & \frac{1}{2}(1 + \frac{K}{4}) = \frac{1}{2}(1 + E)
\end{eqnarray}
where $K = \langle 00\rangle + \langle 01\rangle + \langle 10\rangle - \langle 11\rangle$ is the Clauser-Horne-Shimony-Holt (CHSH) correlation. 

Bell's locality argument in the Clauser-Horne-Shimony-Holt version \cite{CHSH} shows that if Alice and Bob are limited to classical resources, i.e., if they are required to reproduce the correlations on the basis of shared randomness or common causes established before they separate (after which no communication is allowed), then $|K_{C}| \leq 2$, i.e., $|E| \leq 
\frac{1}{2}$, so  the optimal probability of successfully simulating a PR-box is $\frac{1}{2}(1+\frac{1}{2}) = \frac{3}{4}$. 

If Alice and Bob are allowed to base their strategy on shared entangled states prepared before they separate, then the Tsirelson bound for quantum correlations requires that $|K_{Q}| \leq 2\sqrt{2}$, i.e., $|E| \leq \frac{1}{\sqrt{2}}$, so the optimal probability of successful simulation limited by quantum resources is $\frac{1}{2}(1+\frac{1}{\sqrt{2}}) \approx .85$. 

Clearly, the `no signaling' constraint (or relativistic causality) does not rule out simulating a PR-box with a probability greater than $\frac{1}{2}(1+\frac{1}{\sqrt{2}})$. As Popescu and Rohrlich observe, there are possible worlds described by  `superquantum' theories that allow nonlocal boxes with `no signaling' correlations  stronger than quantum correlations, in the sense that $\frac{1}{\sqrt{2}} \leq E \leq 1$. The correlations of a PR-box saturate the CHSH inequality ($E=1$), and so represent a limiting case of `no signaling' correlations.

We see now that Wheeler's question `why the quantum' has two aspects: why is the world quantum and not classical, and why is it quantum rather than superquantum, i.e., \emph{why the Tsirelson bound?} In the following section, I discuss a remarkable answer to this question proposed by Paw\l owski et al \cite{Pawlowski+2009}.

\section{Information Causality}

Paw\l owski et al \cite{Pawlowski+2009} consider a condition they call `information causality,' that the information gain for Bob about an unknown data set of Alice, given all his local resources and $m$ classical bits communicated by Alice, is at most $m$ bits.\footnote{The restriction to the communication of classical bits is essential here. Recall that entanglement correlations can be exploited to allow Alice to send Bob two classical bits by communicating just one quantum bit.} They remark that the `no-signaling' condition is just information causality for $m=0$: if Alice communicates nothing to Bob, then there is no information in the statistics of Bob's outputs about Alice's data set. Paw\l owski et al show that the Tsirelson bound, $|E| \leq \frac{1}{\sqrt{2}}$, follows from this condition.

To see how they arrive at this startling result, it is convenient to consider the following game (related to oblivious transfer and communication complexity problems; see \cite{vanDamthesis,vanDam2005,Brassard+2006} and Section 4): At each round of the game, Alice receives $N$ random and independent bits $\vec{a}=(a_{0},a_{1}, \ldots, a_{N-1})$. Bob, separated from Alice, receives a value of a random uniformly distributed variable $b\in\{0,2,\ldots,N-1\}$. Alice can send one classical bit to Bob with the help of which Bob is required to guess the value of the $b$-th bit in Alice's list, $a_{b}$, for some value of $b \in \{0,\ldots , N-1\}$. We assume that Alice and Bob are allowed to communicate and plan a mutual strategy before the game starts, but once the game starts the only communication between them is the one classical bit that Alice is allowed to send to Bob at each round of the game. They win a round if Bob correctly guesses the $b$-th bit for the round. They win the game if Bob always guesses correctly over any succession of rounds. Note that Alice must decide on the bit she sends to Bob at each round of the game independently of the value of $b$, which is given to Bob at each round and is unknown to Alice.

Clearly,  Bob will be able to correctly guess the value of one of Alice's bits, assuming they agree in advance about the index $k$ of the bit Alice sends at each round, but Bob's guess will be at chance when the value of $b \neq k$. 

Now, suppose Alice and Bob are equipped with a supply of shared PR-boxes. Paw\l owski et al show that there is a strategy that will allow Alice and Bob to win the game, i.e., for any round, and for any $b\in\{0,2,\ldots,N-1\}$, Bob will be able to correctly guess the value of any designated bit $a_{k}$ in Alice's list $a_{0},a_{1},...,a_{N-1}$. 

Consider first the simplest case $N = 2$, where Alice receives two bits, $a_{0}, a_{1}$. The strategy in this case involves a single shared  PR-box. Alice inputs $a_{0}\oplus a_{1}$ into her part of the box (i.e., $a = a_{0}\oplus a_{1}$) and obtains the output $A$. She sends the bit $x = a_{0}\oplus A$ to Bob. Bob inputs the value of $b$, i.e., 0 or 1, into his part of the box and obtains the output $B$. He guesses $a_{b} = x\oplus B = a_{0}\oplus A\oplus B$. 

Now, the box functions in such a way that $A\oplus B = a\cdot b = (a_{0}\oplus a_{1})\cdot b$. So Bob's guess is $x\oplus B = a_{0}\oplus A\oplus B =  a_{0}\oplus  ((a_{0}\oplus a_{1})\cdot b)$. It follows that if $b=0$, Bob correctly guesses $a_{0}$, and if $b=1$, Bob correctly guesses $a_{0}\oplus  a_{0}\oplus a_{1} = a_{1}$.

Suppose Alice receives four bits, $a_{0}, a_{1}, a_{2}, a_{3}$ ($N=4$). Bob's random variable labeling the bit he has to guess takes four values, $b = 0, 1, 2, 3$, and can be specified by two bits, $b_{0}, b_{1}$: 
\[
b = b_{0}2^{0} + b_{1}2^{1} = b_{0} + 2b_{1}
\] 

The strategy in this case involves an inverted pyramid of PR-boxes: two shared PR-boxes, $L$ and $R$, at the first stage, and one shared PR-box at the final second stage. Alice inputs $a_{0}\oplus a_{1}$ into the $L$ box, and $a_{2}\oplus a_{3}$ into the $R$ box.  Bob inputs $b_{0}$ into both the $L$ and $R$ boxes  and obtains the output $B_{0}$ (the input to one of these boxes will be irrelevant, depending on what bit Bob is required to guess; see below).  At the second stage, Alice inputs $(a_{0}\oplus A_{L})\oplus(a_{2}\oplus A_{R})$ into the shared PR-box, where $A_{L}$ is the Alice-output of the $L$ box and $A_{R}$ is the Alice-output of the $R$ box, and obtains the output $A$. Bob inputs $b_{1}$ into this box and obtains the output $B_{1}$.  Alice then sends Bob the bit $x = a_{0}\oplus A_{L}\oplus A$.

Now, Bob could correctly guess either $a_{0}\oplus A_{L}$ or $a_{2}\oplus A_{R}$, using the elementary $N = 1$ strategy, as $x\oplus B_{1} = a_{0}\oplus A_{L}\oplus A\oplus B_{1}$. Here $A\oplus B_{1} = (a_{0}\oplus A_{L}\oplus a_{2}\oplus A_{R})\cdot b_{1}$. If $b_{1} = 0$, Bob would guess $a_{0}\oplus A_{L}$. If $b_{1} = 1$, Bob would guess $a_{2}\oplus A_{R}$. 

So if Bob is required to guess the value of $a_{0}$ (i.e., $b_{0} = 0, b_{1} = 0$) or $a_{1}$ (i.e., $b_{0} = 1, b_{1} = 0$)--- the input to the PR-box $L$---he guesses $a_{0} \oplus A_{L} \oplus A \oplus B_{1} \oplus B_{0}$, where $B_{0}$ is the Bob-output \emph{of the $L$ box.} Then:
\begin{eqnarray}
a_{0} \oplus A_{L} \oplus A \oplus B_{1} \oplus B_{0} & = & a_{0} \oplus A_{L} \oplus B_{0} \nonumber \\
& = & a_{0} \oplus (a_{0} \oplus a_{1})\cdot b_{0} 
\end{eqnarray}
If $b_{0} =0$, Bob correctly guesses $a_{0}$; if $b_{0} =1$, Bob correctly guesses $a_{1}$.

If Bob is required to guess the value of $a_{2}$ (i.e., $b_{0} = 0, b_{1} = 1$) or $a_{3}$ (i.e., $b_{0} = 1, b_{1} = 1$)--- the input to the PR-box $R$---he guesses $a_{0} \oplus A_{L} \oplus A \oplus B_{1} \oplus B_{0}$, where $B_{0}$ is the Bob-output \emph{of the $R$ box.} Then:
\begin{eqnarray}
a_{0} \oplus A_{L} \oplus A \oplus B_{1} \oplus B_{0} & = & a_{2} \oplus A_{R} \oplus B_{0} \nonumber \\
& = & a_{2} \oplus (a_{2} \oplus a_{3})\cdot b_{0} 
\end{eqnarray}
If $b_{0} =0$, Bob correctly  guesses $a_{2}$; if $b_{0} =1$, Bob correctly guesses $a_{3}$.

These strategies are winning strategies for $N=2$, and $N=4$ (the game for $N=1$ is trivial). Clearly, the strategy for $N=4$ is also a strategy for $N=3$ (there is just one less value of $b$ that Bob has to worry about). By adding more stages (levels) to the inverted pyramid, one obtains a strategy for $N=8$ (four shared PR-boxes at the first stage, two shared PR-boxes at the next stage, and one shared PR-box at the third and final stage), and so on. This is also a strategy for $4 \leq N < 8$, so there is a strategy for any $N$.

The game can be modified to allow Alice to send $m$ classical bits of information to Bob at each round, in which case Bob is required to guess the values of any set of $m$ bits in Alice's list of $N$ bits. In this case, Alice and Bob simply apply the above strategy for any $N$ with $m$ inverted pyramids of PR-boxes, one for each bit in the set of bits Bob is required to guess. 

We have seen that Alice and Bob can win this game if they share PR-boxes ($E=1$). What if they share non-signaling (NS) boxes with \emph{any} `no signaling' correlations corresponding to $|E| < 1$, such as classical correlations($|E| \leq \frac{1}{2}$), or the correlations of entangled quantum states ($|E| \leq \frac{1}{\sqrt{2}}$), or superquantum `no signaling' correlations ($\frac{1}{\sqrt{2}} < E < 1$)? 

The probability of simulating a PR-box with a NS-box is $\frac{1}{2}(1+E)$, where $E$ depends on the NS-box (the nature of the correlations). Consider the $N=4$ game where Alice and Bob share NS-boxes, and Alice is allowed to communicate one bit to Bob. Bob's guess $x \oplus B_{1} \oplus B_{0}$ will be correct if $B_{1}$ and $B_{0}$ are both correct or both incorrect (since  $B_{1} \oplus B_{0}$ will be the same in either case).

The  probability of being correct at both stages is:
\begin{equation}
\frac{1}{2}(1+E)\cdot \frac{1}{2}(1+E) = \frac{1}{4}(1+E)^{2}
\end{equation}
The  probability of being incorrect at both stages is:
\begin{equation}
(1-\frac{1}{2}(1+E))\cdot(1-\frac{1}{2}(1+E)) = \frac{1}{2}(1-E)\cdot \frac{1}{2}(1-E) = \frac{1}{4}(1-E)^{2}
\end{equation}
So the probability $P_{k}$ that Bob guesses correctly, i.e., the probability that $\beta = a_{k}$ when $b=k$, is:
\begin{equation}
P_{k} = \frac{1}{4}(1+E)^{2}+\frac{1}{4}(1-E)^{2} = \frac{1}{2}(1+E^{2})
\end{equation}

In the general case $N=2^{n}$, Bob guesses correctly if he makes an even number of errors over the $n$ stages ($B_{0}, B_{1}, B_{2}, \ldots$) and the probability is:
\begin{equation}
P_{k} = \frac{1}{2^{n}}(1+E)^{n}    +  \frac{1}{2^{n}} \sum_{j=1}^{\lfloor\frac{n}{2}\rfloor} {n \choose 2 j} (1-E)^{2j}  (1+E)^{n-2j}         = \frac{1}{2}(1+E^{n})
\end{equation}
where $\lfloor \frac{n}{2} \rfloor$ denotes the integer value of $\frac{n}{2}$.
For example, if $n=3$, the probability of being correct at each stage is:
\begin{equation}
\frac{1}{2}(1+E)\cdot \frac{1}{2}(1+E)\cdot \frac{1}{2}(1+E)
\end{equation}
and the probability of being incorrect at two out of the three stages (i.e., at $B_{0}, B_{1}$ or $B_{0}, B_{2}$ or $B_{1}, B_{2}$ is:
\begin{equation}
3\cdot \frac{1}{2}(1-E)\cdot \frac{1}{2}(1-E)\cdot \frac{1}{2}(1+E)
\end{equation}
so the probability that Bob guesses correctly is :
\begin{equation}
P_{k} = \frac{1}{8}(1+E)^{3} + \frac{3}{8}(1-E)^{2}(1+E) = \frac{1}{2}(1+E^{3})
\end{equation}

\section{The Tsirelson bound}

In the game considered above, Alice has a list of $N$ bits and Bob has to guess an arbitrarily selected one of these bits, $b=k$. If Bob knows the value of the bit he has to guess, $P_{k} = 1$. The binary entropy of $P_{k}$ is defined as $h(P_{k}) = -P_{k}\log P_{k} - (1-P_{k})\log (1-P_{k})$, so $h(P_{k}) = 0$. If Bob has no information about the bit he has to guess, $P_{k} = 1/2$, i.e., his guess is at chance, and $h(P_{k}) = 1$. 

If Alice sends Bob one classical bit of information, information causality requires that Bob's information about the $N$ unknown bits increases by at most one bit.  So if the bits in Alice's list are unbiased and independently distributed, Bob's  information about an arbitrary bit $b=k$ in the list cannot increase by more than $1/N$ bits, i.e., for Bob's guess about an arbitrary bit in Alice's list, the binary entropy $h(P_{k})$ is at most $1/N$ closer to 0 from the chance value 1, i.e., $h(P_{k}) \geq 1-1/N$.

It follows that the condition for a violation of information causality in this case can be expressed as:  
\begin{equation}
h(P_{k}) < 1 - 1/N \label{eqn:icviolate1}
\end{equation}
or, taking $N = 2^{n}$, the condition is: 
\begin{equation}
h(P_{k}) < 1-\frac{1}{2^{n}} \label{eqn:icviolate2}
\end{equation}
Since $P_{k} =\frac{1}{2}(1+E^{n})$,   we have a violation of information causality when:
\begin{equation}
h(\frac{1}{2}(1+E^{n}))  < 1-\frac{1}{2^{n}} \label{eqn:violation1}
\end{equation}

Paw\l owski et al \cite{Pawlowski+2009} make use of the following inequality:
\begin{equation}
h(\frac{1}{2}(1+y)) \leq 1-\frac{y^{2}}{2 \ln{2}} \label{eqn:inequal}
\end{equation}
where $\ln 2 \approx .693$ is the natural log of 2 (base $e$). So  information causality is violated if
\begin{equation}
1- \frac{E^{2n}}{2\ln 2} < 1 -  \frac{1}{2^{n}}
\end{equation}
i.e., if
\begin{equation}
(2E^{2})^{n} > 2 \ln 2 \approx 1.386 \label{eqn:violation2}
\end{equation}

If $2E^{2} = 1$, i.e., if $E = E_{T} = \frac{1}{\sqrt{2}}$ (the Tsirelson bound), the inequality (\ref{eqn:violation2}) is satisfied. This is a sufficient condition for a violation of information causality, but it is not necessary: even if $(2E_{T}^{2})^{n} \not > 2 \ln 2$, we could still have a violation of information causality for some $n$ if $h(\frac{1}{2}(1+E_{T}^{n})) < 1-\frac{1}{2^{n}}$. See the Appendix for a proof that information causality is satisfied for $E=E_{T}$, i.e.,  $h(\frac{1}{2}(1+E_{T}^{n})) \geq 1-\frac{1}{2^{n}}$ for any $n$.

If $E > E_{T}$, i.e., if $2E^{2} = 1+a$, for some $a$, no matter how small, there is a violation: $(2E^{2})^{n}  >  1 + na$,\footnote{Recall that $(1+a)^{n}$ can be expanded as $(1+a)^{n}  = 1 + na  +  \frac{n(n-1)}{2!}+ \frac{n(n-1)(n-2)}{3!} + \cdots$} but $1+na > 2 \ln 2 \approx 1.386$ for some $n$. That is, for any $a$, however small, there is a value of $n$ such that $n > \frac{.386}{a}$, hence a value of $n$ for which information causality is violated.

To appreciate the significance of this result, consider some numbers for $E$  and $n$.  The condition for a violation of information causality is $h(P_{k})  < 1-\frac{1}{2^{n}}$. Recall that $\log_{2} x = \frac{\log_{10} x}{\log_{10} 2} \approx \frac{\log_{10} x}{.301}$.

Consider first the case where $E = E_{T} = \frac{1}{\sqrt{2}} \approx .707$, theTsirelson bound.

When $n = 1$,  Alice has $2^{1}=2$ bits:
\begin{eqnarray}
h(P_{k})  & = & -(\frac{1}{2}(1+\frac{1}{\sqrt{2}}) \frac{\log_{10} \frac{1}{2}(1+\frac{1}{\sqrt{2}})}{.301} \nonumber \\
&& + \frac{1}{2}(1-\frac{1}{\sqrt{2}}) \frac{\log_{10} \frac{1}{2}(1-\frac{1}{\sqrt{2}})}{.301}) \nonumber \\
& \approx & .600 \label{eqn:n=1}
\end{eqnarray}
There is no violation of information causality because $.600 > 1 - \frac{1}{2^{1}} = \frac{1}{2}$.

When $n = 10$, Alice has $2^{10}=1024$ bits:
\begin{eqnarray}
h(P_{k}) &=& -(\frac{1}{2}(1+\frac{1}{\sqrt{2}^{10}}) \frac{\log_{10} \frac{1}{2}(1+\frac{1}{\sqrt{2}^{10}})}{.301} \nonumber \\
&& + \frac{1}{2}(1-\frac{1}{\sqrt{2}^{10}}) \frac{\log_{10} \frac{1}{2}(1-\frac{1}{\sqrt{2}^{10}})}{.301}) 
\approx  .99939
\end{eqnarray}
There is still no violation of information causality because  $.99939 > 1 - \frac{1}{2^{10}} = 1 - \frac{1}{1024} \approx .9990$.

Now consider the case where $E > E_{T}$. Take $E = .725$ and $n = 7$. In this case, there is a violation of information causality: 
\begin{eqnarray}
h(P_{k}) &=& -(\frac{1}{2}(1+ .725^{7}) \frac{\log_{10} \frac{1}{2}(1+ .725^{7})}{.301} \nonumber \\
&& + \frac{1}{2}(1- .725^{7}) \frac{\log_{10} \frac{1}{2}(1- .725^{7})}{.301})\nonumber \\
& \approx & .99208
\end{eqnarray}
There is a violation of information causality because $.99208 < 1 - \frac{1}{128} \approx .99218$. There is no violation for $n=6$ because $.9848 > 1 - \frac{1}{64} \approx .9844$.

Note that the inequality (\ref{eqn:inequal}) has not been used in the above calculations. The only role of the inequality is to allow one to easily see that information causality is violated for \emph{some} value of $n$ if $E > E_{T}$, i.e., if $2E^{2} > 1 + a$ for any $a$. In fact, information causality could be violated for a lower value of $n$. In the case above, $E = .725$, $a \approx .05125$. Using the inequality, we find that information causality is violated when $n > \frac{.386}{a}$, i.e., when $n \geq 8$.

 If $E$ is very close to the Tsirelson bound, then $n$ must be very large for a violation of information causality. For $n=10$ and $E = .708$:
\begin{eqnarray}
h(P_{k}) &=& -(\frac{1}{2}(1+ .708^{10}) \frac{\log_{10} \frac{1}{2}(1+ .708^{10})}{.301} \nonumber \\
&& + \frac{1}{2}(1- .708^{10}) \frac{\log_{10} \frac{1}{2}(1- .708^{10})}{.301}) 
 \approx  .99938
\end{eqnarray}
There is no violation of information causality because $.99938 > 1 - \frac{1}{1024} \approx .9990$. Using the inequality, with $a = .708 - \frac{1}{\sqrt{2}}$, we find that $n \geq 432$ for a violation of information causality.

Another way to look at this:
If  $E = E_{T} = \frac{1}{\sqrt{2}}$, $P_{k} = \frac{1}{2}(1+E^{n}) \rightarrow \frac{1}{2}$ and $h(P_{k})  \rightarrow  1$  as $n \rightarrow \infty$. So, if Alice has a very long list and sends Bob one bit of information, Bob's ability to correctly guess an arbitrary bit in Alice's list is essentially at chance if the correlations are bounded by the Tsirelson bound. For a PR-box, $E = 1$, $P_{k} = 1$, $h(P_{k}) = 0$, so Bob can correctly guess any arbitrary bit in Alice's list.

\section{Comments}

The analysis in Section 3 related information causality directly to a condition on the binary entropy. In Paw\l owski et al \cite{Pawlowski+2009}, the authors relate information causality directly to a condition on the mutual information between Alice and Bob, and only indirectly to the binary entropy:
\begin{quotation}
Ideally, we wish to define that information causality holds if, after transfer of the $m$-bit message, the mutual information between AliceÕs data $\vec{a}$ and everything that Bob has---that is, the message $\vec{x}$ and his part $B$ of the previously shared correlation---is bounded by $m$. Intuitively appealing though such a definition is, it has the severe issue that it is not theory-independent. Specifically, a mutual information expression `$I(\vec{a}:\vec{x},B)$' has to be defined for a state involving objects from the underlying theory (the possibilities include classical correlation, a shared quantum state and NS-boxes). It is far from clear whether mutual information can be defined consistently for all nonlocal correlations, nor whether such a definition would be unique.
\end{quotation}

Paw\l owski et al  denote Bob's output by $\beta$ and quantify the efficiency of Alice's and Bob's strategy by:
\begin{equation}
  I \equiv \sum_{k=0}^{N-1} I(a_k : \beta | b = k) 
\end{equation}
where $I(a_k: \beta | b = k)$ is the Shannon mutual information between $a_k$ and $\beta$, computed under the condition that Bob is required to guess the bit $b=k$. 
They show that if the mutual information $I(\vec{a}:\vec{x},B)$ for any `no signaling' theory satisfies three  constraints (which are satisfed for quantum information and for classical information, a special case of quantum information):
\begin{itemize}
\item consistency with the classical Shannon mutual information when the Alice and Bob subsystems are both classical
\item the data-processing inequality: any local manipulation of data can only degrade information, i.e., acting on one subsystem locally by any admissible transformation cannot increase the mutual information
\item the chain rule:
$
I(A:B,C) = I(A:C) + I(A:B|C)
$,
where $I(A:B|C)$ is the conditional mutual information
\end{itemize}
then (i) information causality is satisfied, i.e., 
$
I(\vec{a}:\vec{x},B) \leq m
$,
and (ii)
$
I(\vec{a}:\vec{x},B) \geq I
$. 

Since $I(\vec{a}:\vec{x},B) > m$ if $I > m$,  it follows  that information causality is violated if: 
\begin{equation}
I > m \label{eqn:icviol}
\end{equation}
So if information causality is satisfied, then $I \leq m$, i.e., $I \leq m$ is a necessary condition for information causality. (Note that we could, of course, have $I \leq m$ but $I(\vec{a}:\vec{x},B) > m$, so (\ref{eqn:icviol}) is not a sufficient condition for information causality.) As the authors emphasize, $I$ is fully specified by Alice's and Bob's input and output bits and is independent of the details of any particular physical theory.

The Shannon mutual information $I(X\!:\!Y)$ of two random variables is a measure of how much information they have in common: the sum of the information content of the two random variables, as measured by the Shannon entropy (in which joint information is counted twice), minus their joint information:
\begin{eqnarray}
I(X\!:\!Y) &  = & H(X) + H(Y) - H(X,Y) \nonumber\\
& = & H(X) - H(X|Y)
\end{eqnarray}
where $H(X) =  -\sum_{i}p_{i}\log p_{i}$ is the Shannon entropy of the random variable $X$, $H(X,Y) =  -\sum_{i,j}p_{i,j}\log p_{i,j}$ is the  joint Shannon entropy of the two random variables $X,Y$ representing the joint information, and $H(X|Y)$ is the conditional entropy: $H(X|Y) = H(X,Y) - H(Y)$. Note that $H(X|Y) \leq H(X)$, with equality if and only if $X,Y$ are independent.

So: 
\begin{equation}
  I \equiv \sum_{k=0}^{N-1} I(a_k : \beta | b = k) = \sum_{k=0}^{N-1} (H(a_{k}) + H(\beta) - H(a_{k},\beta))
\end{equation}
where the condition $b=k$ has been omitted for ease of reading.

First note that
\begin{eqnarray}
H(a_{k}|\beta) & = & H(a_{k}\oplus \beta|\beta) \nonumber \\
& \leq & H(a_{k}\oplus\beta)
\end{eqnarray}
The first equality follows because only the probabilities of the different alternatives are relevant in the calculation of the entropy. In this case, the probabilities are 0 and 1 and, given that $\beta = 0$, the probability that $a_{k} = 0$ is the same as the probability that $a_{k}\oplus \beta = 0$, i.e., that $a_{k} = \beta$, and the probability that $a_{k} = 1$ is the same as the probability that $a_{k}\oplus \beta = 1$, i.e., that $a_{k} \neq \beta$; and similarly if $\beta = 1$. The second inequality follows because conditioning decreases entropy.

Now:
\begin{equation}
 H(a_{k}\oplus\beta) = h(P_{k})
 \end{equation}
 so
 \begin{equation}
 H(a_{k}|\beta) \leq h(P_{k})
 \end{equation}
 It follows that:
 \begin{equation}
 I(a_{k}:\beta)|b=k) \geq H(a_{k}) - h(P_{k})
 \end{equation}
 In the case where the bits in Alice's list are unbiased and independently distributed, $H(a_{k}) = 1$, so:
 \begin{equation}
I(a_{k}:\beta)|b=k) \geq 1- h(P_{k})
 \end{equation}
 i.e.,
 \begin{equation}
 I \geq N - \sum_{k=0}^{N-1}h(P_{k})
 \end{equation}
and since $h(P_{k}) = \frac{1}{2}(1+E^{n})$, which is independent of $k$:
\begin{equation}
 I \geq N - Nh(P_{k})
\end{equation}

For a PR-box, $E=1$, $h(P_{k}) = 0$, and $I = N$.  If Bob guesses randomly for all $k$, then $h(P_{k}) = 1$, $I = 0$. So in the case where Alice sends $m$ bits of information to Bob, $0 \leq I \leq N$, with a violation of information causality when $I > m$.

If Alice sends Bob one bit of information, information causality is violated if $I > 1$, i.e., if:
\begin{equation}
h(P_{k}) < 1-\frac{1}{N}
\end{equation}
or, taking $N = 2^{n}$, if:
\begin{equation}
h(P_{k}) < 1-\frac{1}{2^{n}}
\end{equation}
which are, respectively, equations (\ref{eqn:icviolate1}) and (\ref{eqn:icviolate2}) of Section 3.

Paw\l owski et al \cite[p. 1101]{Pawlowski+2009} express the condition of information causality as follows:
\begin{quotation}
Formulated as a principle, information causality states: `the information gain that Bob can reach about a previously unknown to him data set of Alice, by using all his local resources and $m$ classical bits communicated by Alice, is at most $m$ bits.' The standard no- signalling condition is just information causality for $m = 0$. 
\end{quotation}

Stated in this way, the condition seems trivial: of course, if Alice sends Bob $m$ bits of information, his information gain is at most $m$ bits, and if $m=0$ his information gain is 0. But implicit in the condition is that Bob's local resources include the marginal probabilities of correlations between Alice and Bob and the values of the correlated variables, and similarly for Alice. The issue concerns the extent to which Alice and Bob can exploit previously established correlations between them in such a way that the $m$ bits of information communicated by Alice to Bob will allow Bob to correctly guess an arbitrarily designated set of bits in Alice's data set, which might contain $N > m$ bits. Of course, without exploiting the correlations, Bob can know some specific, previously agreed upon set of $m$ bits and, exploiting classical correlations, i.e., previously established shared randomness, Bob can know a different specific set of $m$ bits on each occasion that Alice sends him $m$ bits.\footnote{A suitably long shared list of random bits can be used by Alice and Bob to pick a different set of $m$ bits at each round of the guessing game, for some finite set of rounds.} The relevant insight is that if the correlations are PR-box correlations, then Alice can send Bob a  set of $m$ bits chosen on the basis of the Alice-values of the correlated variables, where Alice and Bob select the variables appropriately as the inputs to the PR-boxes,  in such a way that   Bob can  correctly guess \emph{any arbitrary set of $m$ bits in Alice's data set}. In other words, for the case $m=1$, there is a way of exploiting the PR-box correlations so that the one bit of information can be associated with \emph{any designated bit} in Alice's data set of $N$ bits, for any $N$ (this was pointed out already in \cite{Wolf+2005}). 

So in the case where the bits in Alice's data set are unbiased and independently distributed and Alice sends Bob one bit of information, the PR-box correlations can be exploited to achieve $P_{k} = 1$ for all $k$, i.e., $h(P_{k}) = 0$ for all $k$. The intuition behind information causality is that this is `too good to be true,' in fact, that the binary entropy should be bounded: $h(P_{k}) \geq 1 - \frac{1}{N}$. Putting it differently, when the bits in Alice's data set are unbiased and independently distributed, the intuition is that if the correlations can be exploited to distribute one bit of communicated information among the $N$ unknown bits in Alice's data set, the amount of information distributed should be no more than $\frac{1}{N}$ bits, because there can be no information about the bits in Alice's data set in the previously established correlations themselves. 

As Paw\l owski et al show, for `no signaling' correlations, $P_{k} = \frac{1}{2}(1+ E^{n})$, where $N = 2^{n}$. For classical correlations, $E = \frac{1}{2}$, $h(P_{k}) \approx .811$ for $n=1$. For quantum correlations, $E = E_{T} = \frac{1}{\sqrt{2}}$, $h(P_{k}) \approx .600$ for $n=1$, so Alice and Bob can do better exploiting quantum correlations than they can if they are restricted to classical correlations. This is the case for any $n$, but information causality is always satisfied. The intriguing result by  Paw\l owski et al is that information causality is violated \emph{ for some value of $n$} if $E > E_{T}$. From this perspective, it is misleading to claim that the `no signaling' condition is `just information causality for $m=0$.' If Alice communicates no information to Bob, they have no possibility of exploiting  correlations to increase Bob's access to Alice's data set. The condition of information causality concerns the extent to which correlations can be exploited to increase Bob's access to Alice's data set, in the sense of improving Bob's ability to correctly guess any arbitrary bit in Alice's data set.

In fact, the term `information causality' is suggestive in the wrong sense. The principle really has nothing to do with causality and is better understood as \emph{a constraint on the ability of correlations to enhance the information content of communication in a distributed task}. A more appropriate term would be `informational neutrality of correlations,' and the principle should be formulated as follows:
\begin{quotation}
Correlations are informationally neutral: insofar as they can be exploited to allow Bob to distribute information communicated by Alice among the bits in an unknown data set held by Alice in such a way as to increase Bob's ability to correctly guess an arbitrary bit in the data set, they cannot increase Bob's information about  the data set  by more than the number of bits communicated by Alice to Bob. 
\end{quotation}
So if Alice has a data set of $N$ uniformly and independently distributed bits and sends Bob one bit of information, and Bob can exploit previously established correlations to increase his ability to correctly guess an arbitrary bit in the data set, his information gain about an arbitrary bit in the data set can be  no more than $1/N$ bits, i.e., the binary entropy of the probability of a correct guess cannot be less than $1 - 1/N$.

The correlations of a PR-box are not informationally neutral in this sense. While they are logically admissible, they are `too good to be true' in the way they allow the solution of the following two distributed tasks:

\begin{itemize}
\item 
The `dating game': Alice and Bob would like to go on a date, but only if they know that they both like each other. In other words, they would like to compute a function that takes the value 1 if they both like each other (i.e., if both inputs to the function are 1), but takes the value 0 if at least one party does not like the other (i.e., if the inputs are both 0, or one input is 0 and the other input is 1). Now, in the real world, there is no way they can do this without revealing information that they both want to keep private: Alice does not want Bob to know that she likes him \emph{if he does not like her}, and similarly for Bob. With a PR-box, they can compute this function, while keeping private the information they want to keep private. Alice and Bob input 0 or 1 into their inputs to the PR-box when they are separate (so neither party sees the other's input). They then come together and share the outputs. If the outputs are different, they know that both inputs were 1, so they happily go on a date. In this case, of course, Alice knows that Bob likes her, and Bob knows that Alice likes him, but that's fine. If the outputs are the same, they know only that either Alice did not like Bob, or that Bob did not like Alice, or that the dislike was mutual. While Alice can infer that Bob does not like her if she likes him, this knowledge is private, so Alice avoids any humiliation; and similarly for Bob.
\item `One-out-of-two' oblivious transfer: Alice has a data set consisting of two bits of information. The constraint on Alice is that she can send Bob one bit of information. The requirement for Bob is that he uses the one bit of communicated information to correctly guess whichever bit he chooses in Alice's data set, in such a way that Alice is oblivious of his choice. Again, there is no way to do this in the real world, but if Alice and Bob have access to a PR-box they can successfully achieve this task. The protocol is the same as the protocol for the $N=2$ case discussed in Section 2.
\end{itemize}

The remarkable result of Paw\l owski et al shows that, while quantum correlations are `more like' PR-box correlations than classical correlations, insofar as they increase the ability of Alice and Bob to perform distributed tasks relative to classical correlations, they represent the limit of what is possible if correlations are `informationally neutral,' in the sense that correlations established prior to the choice of a data set can contain no information about such a data set, and hence should not be able to be exploited to allow a party who has no access to the data set to correctly guess any arbitrary bit in the set. This considerably extends related results by van Dam \cite{vanDamthesis,vanDam2005}, Brassard et al \cite{Brassard+2006}, Linden et al \cite{Linden+2006}. Note that there are other results in which nonlocal boxes are exploited to derive the Tsirelson bound. See Skrzypczyk et al \cite{Skrzypczyk+2008}, in which a dynamics is defined for PR-boxes and the Tsirelson bound is derived from a condition called `nonlocality swapping.'

Paw\l owski et al \cite[p.1103--1104]{Pawlowski+2009} conclude with the following remarks:
\begin{quotation}
In conclusion, we have identified the principle of Information Causality, which precisely distinguishes physically realized correlations from nonphysical ones (in the sense that quantum mechanics cannot reach them). It is phrased in operational terms and in a theory-independent way and therefore we suggest it is at the same foundational level as the no-signaling condition itself, of which it is a generalization.

The new principle is respected by all correlations accessible with quantum physics while it excludes all no-signaling correlations, which violate the quantum Tsirelson bound.
Among the correlations that do not violate that bound it is not known whether Information Causality singles out exactly those allowed by quantum physics.
If it does, the new principle would acquire even stronger status.
\end{quotation}

Classical correlations bounded by $E \leq \frac{1}{2}$ can be associated with a polytope, where the vertices represent `no signaling' deterministic states. For example, in the case considered above for a bipartite system with  two binary-valued quantities, the deterministic state in which the values of the two quantities are both zero, for all four possible combinations, is given by Table 2.
 \begin{table}[h!]
\begin{center}
\begin{tabular}{|ll||ll|ll|} \hline
   &$a$&$0$ & &$1$&\\
   $b$&&&&&\\\hline\hline
  $0$ &&$p(00|00) = 1$&$ p(10|00) = 0$  & $p(00|10) = 1$&$ p(10|10) = 0$     \\
   &&$p(01|00) = 0$&$p(11|00) = 0$  & $p(01|10)=0$&$ p(11|10) = 0$  \\\hline
   $1$&&$p(00|01)=1$&$ p(10|01)=0$  & $p(00|11)=1$&$ p(10|11)=0$   \\
  &&$p(01|01)=0$&$ p(11|01)=0$  & $p(01|11)=0$&$ p(11|11)=0$   \\\hline
\end{tabular}
\end{center}
 \caption{A deterministic state}
\end{table}
There are 16 `no signaling' deterministic states (each of which can be represented as a product of local states, an Alice deterministic state and a Bob deterministic state) out of 256 possible deterministic states---the remaining 240 deterministic states allow signaling. The 16-vertex classical polytope is included in a 24-vertex `no signaling' nonlocal polytope, where the vertices are the 16 `no signaling' deterministic states and 8 additional PR-box states, represented by the probabilities in Table 1, or probabilities obtained from Table 1 by by relabeling the $a$-inputs, and the $A$-outputs conditionally on the $a$-inputs, and the $b$-inputs, and the  $B$-outputs conditionally on the $b$-inputs. Quantum correlations bounded by $E = E_{T} \leq \frac{1}{\sqrt{2}}$ are associated with a spherical convex set with extremal points between the 16-vertex classical simplex and the 24-vertex `no signaling' nonlocal polytope. 

The open question is whether non-quantum correlations  represented by points outside the quantum convex set but below the Tsirelson bound can also be excluded by information causality. For a discussion, see Allcock et al \cite{Allcock+2009}. 

\section{Appendix}

In \cite{Pawlowski+2009}, the authors prove quite generally that information causality is satisfied for any `no signaling' theory satisfying three constraints on mutual information (consistency with the classical Shannon mutual information, the data-processing inequality, and the chain rule), hence for quantum information, which satisfies the constraints. It follows that information causality is satisfied at and below the the Tsirelson bound. 

The following is a simple direct proof (see Section 3) that  if $E = E_{T} = \frac{1}{\sqrt{2}}$, then:
\begin{equation}
h(\frac{1}{2}(1 + E^{n}) \geq 1 - \frac{1}{2^{n}} 
\end{equation}
i.e., 
\begin{equation}
-\frac{1}{2}(1+E^{n})\log (\frac{1}{2}(1+E^{n})) -\frac{1}{2}(1-E^{n})\log (\frac{1}{2}(1-E^{n})) \geq 1 - \frac{1}{2^{n}} 
\end{equation}
After a little algebra, this can be expressed as:
\begin{equation}
\log (1-E^{2n}) + E^{n}\log \frac{1+E^{n}}{1-E^{n}} \leq \frac{1}{2^{n-1}} \label{eqn:ic}
\end{equation}
Note that the logarithms are to the base 2.

Now, if $-1 \leq x \leq 1$:
\begin{eqnarray}
\log_{e}(1+x) & = & x -\frac{1}{2}x^{2} + \frac{1}{3}x^{3} - \frac{1}{4}x^{4} +  \cdots \label{eqn:log} \\
\log_{e}\frac{1+x}{1-x} & = & 2(x + \frac{x^{3}}{3} + \frac{x^{5}}{5} \cdots + \frac{1}{2m-1}x^{2m-1} + \cdots)
\end{eqnarray}
So 
\begin{equation}
\log_{e} (1-E^{2n}) + E^{n}\log_{e} \frac{1+E^{n}}{1-E^{n}} = E^{2n} + \frac{1}{6}E^{4n} + \frac{1}{15}E^{6n} \cdots + \frac{1}{m(2m-1)}E^{2mn} + \cdots
\end{equation}
Substituting $E=E_{T}=\frac{1}{\sqrt{2}}$, this becomes:
\begin{equation}
(\frac{1}{2})^{n} + \frac{1}{6}\cdot(\frac{1}{2})^{2n} + \frac{1}{15}\cdot(\frac{1}{2})^{3n} \cdots + \frac{1}{m(2m-1)}\cdot(\frac{1}{2})^{mn} + \cdots
\end{equation}

Since $\log_{2} x = \log_{2}e \cdot \log_{e}x$, it follows that $\log (1-E^{2n}) + E^{n}\log \frac{1+E^{n}}{1-E^{n}}$, where the logarithms are to the base 2, can be expressed as the following infinite series:
\begin{equation}
\log_{2}e \cdot (\frac{1}{2^{n}} + \frac{1}{6}\cdot \frac{1}{2^{2n}} + \frac{1}{15}\cdot \frac{1}{2^{3n}} \cdots + \frac{1}{m(2m-1)}\cdot \frac{1}{2^{mn}} + \cdots)
\end{equation}
so the inequality (\ref{eqn:ic}) we are required to prove becomes:
\begin{equation}
 \frac{1}{2^{n}} + \frac{1}{6}\cdot \frac{1}{2^{2n}} + \frac{1}{15}\cdot \frac{1}{2^{3n}} \cdots + \frac{1}{m(2m-1)}\cdot \frac{1}{2^{mn}} + \cdots \leq \log_{e}2 \cdot \frac{1}{2^{n-1}}
\end{equation}
or
\begin{equation}
\frac{1}{2} + \frac{1}{6}\cdot \frac{1}{2^{n+1}} + \frac{1}{15}\cdot \frac{1}{2^{2n+1}} \cdots + \frac{1}{m(2m-1)}\cdot \frac{1}{2^{(m-1)n+1}} + \cdots \leq \log_{e}2 \approx .693147 \label{eqn:ic2}
\end{equation}
This is clearly the case. The largest value of the series is obtained for $n=1$, when the first term is $.5$.  The remaining terms affect only the second and later decimal places. 

Alternatively, from (\ref{eqn:log}) we have:
\begin{equation}
\log_{e}2 = 1 - \frac{1}{2} + \frac{1}{3} - \frac{1}{4} + \cdots
\end{equation}
so, subtracting the series on the left hand side of the inequality (\ref{eqn:ic2}) from the series for $\log_{e}2$,  what has to be proved is that, for any $n$:
\begin{equation}
(\frac{1}{3} - \frac{1}{6}\cdot \frac{1}{2^{n+1}}) - (\frac{1}{4} + \frac{1}{15}\cdot \frac{1}{2^{2n+1}}) + (\frac{1}{5} - \frac{1}{28}\cdot \frac{1}{2^{3n+1}}) - (\frac{1}{6} + \frac{1}{45}\cdot \frac{1}{2^{4n+1}}) + \cdots \geq 0
\end{equation}
This is obvious by inspection, since each negative term in parenthesis is smaller than its postive predecessor, for any $n$.

\section*{Acknowledgements}
This paper was written during the tenure of a University of Maryland semester RASA award.

\bibliographystyle{plain}
\bibliography{whytsirelson3}

\end{document}